\documentclass[acmlarge, screen, authorversion]{acmart}
\AtBeginDocument{%
  \providecommand\BibTeX{{%
    \normalfont B\kern-0.5em{\scshape i\kern-0.25em b}\kern-0.8em\TeX}}}

\setcopyright{acmcopyright}
\copyrightyear{2023}
\acmYear{2023}
\acmDOI{XXXXXXX.XXXXXXX}

\acmConference[Conference acronym 'XX]{Make sure to enter the correct
  conference title from your rights confirmation emai}{June 03--05,
  2018}{Woodstock, NY}
\acmPrice{15.00}
\acmISBN{978-1-4503-XXXX-X/18/06}

\begin{document}

\title[Co-Designing and Implementing an Accessible Text-to-Image Interface]{Breaking Barriers to Creative Expression: Co-Designing and Implementing an Accessible Text-to-Image Interface}

\author{Atieh Taheri}
\email{a_taheri@ucsb.edu}
\affiliation{
\institution{University of California, Santa Barbara}
\city{Santa Barbara}
\state{California}
\country{USA}
}
\author{Mohammad Izadi}
\email{izadi@google.com}
\affiliation{%
  \institution{Google}
  \city{Mountain View}
  \state{California}
  \country{USA}
}
\email{izadi@google.com}

\author{Gururaj Shriram}
\affiliation{%
  \institution{Google}
  \country{USA}}
\email{gshriram@google.com}

\author{Negar Rostamzadeh}
\affiliation{%
  \institution{Google}
  \city{Montreal}
  \country{Canada}
}
\email{nrostamzadeh@google.com}

\author{Shaun Kane}
\affiliation{%
  \institution{Google}
  \state{Colorado}
  \country{USA}}
\email{shaunkane@google.com}

% \affiliation{%

% \affiliation{%

% \affiliation{%

% \affiliation{%

\renewcommand{\shortauthors}{Taheri et al.}

\begin{abstract}
  Text-to-image generation models have grown in popularity due to their ability to produce high-quality images from a text prompt. One use for this technology is to enable the creation of more accessible art creation software. In this paper, we document the development of an alternative user interface that reduces the typing effort needed to enter image prompts by providing suggestions from a large language model, developed through iterative design and testing within the project team. The results of this testing demonstrate how generative text models can support the accessibility of text-to-image models, enabling users with a range of abilities to create visual art.
\end{abstract}






\maketitle

\section{Introduction}

% \begin{itemize}
%     \item T2I is emerging as a powerful creative tool
%     \item This can be accessible! Movements that may be difficult for some (e.g. drawing straight lines) can be overcome
%     \item UI for these interfaces is still an emerging area (e.g. Bing with Dall-E integrated)
%     \item This paper describes a collaborative effort to identify
% \end{itemize}
Recent advances in text-to-image (T2I) models~\cite{reed2016generative, zhang2017stackgan, hong2018inferring, johnson2018image, zhang2018photographic} have enabled users to create high-quality, stylistically diverse images of nearly any subject with little or no artistic training~\cite{chang2023prompt}. Models such as DALL-E 2~\footnote{\url{https://openai.com/dall-e-2}}, Midjourney~\footnote{\url{https://www.midjourney.com}}, and Stable Diffusion~\footnote{\url{https://stability.ai}} allow users to enter a text or text$+$image prompt, which the system interprets to generate images that themselves are not contained within the training set.

% Recent advances in  deep learning and computer vision technologies~\cite{reed2016generative, zhang2017stackgan, hong2018inferring, johnson2018image, zhang2018photographic} have significantly advanced the capabilities of text-to-image (T2I) models. Current T2I models can create high quality images of many subjects and in many image styles, including both photorealistic images and images in artistic styles such as oil paintings or cartoons. T2I models can be used to support trained artists in their creative work~\cite{chang2023prompt}, and enable inexperienced artists to create high-quality images.

% Current T2I models such as DALL-E 2\footnote{\url{openai.com/product/dall-e-2}} and Midjourney\footnote{\url{www.midjourney.com}} are operated through \emph{prompting}: the user creates a text-based prompt that describes the desired image, including the subject of the image and its style. For example, a creator might enter the prompt \emph{``photograph of an orange cat riding on a skateboard''} to generate a photorealistic image of their intended subject. In many cases, the T2I model will return multiple candidate images for the given prompt; the user can choose one of the images or revise their prompt and try again.

T2I models have gained attention in part because they are enable non-artists to create images that may be of a similar quality to those created by professional artists. However, these models also have potential to help would-be artists who encounter accessibility issues during their artistic work. Specifically, we see potential for T2I models to serve as an accessible image creation method for those with motor disabilities, as it sidesteps the need for fine motor control that might occur when using a paintbrush to paint on canvas or using a mouse to edit pixels in a digital image~\cite{wobbrock2011ability}. Most current T2I models generate images based on a text prompt (\emph{e.g.}, ``photograph of an orange cat riding on a skateboard''), which is ideal for accessibility as text can be entered in many ways, including traditional keyboards, alternative input devices, eye or head movements, or voice~\cite{koester2018text}.

While text input can accommodate users with a range of abilities, text entry may still be challenging for users with motor disabilities~\cite{polacek2017text, mackenzie2002text, mott2017dwell}. Thus, it is important that accessible T2I tools be designed to support user interaction that best suits their abilities~\cite{wobbrock2011ability}, while maintaining the sophistication needed to support users' creative goals. 

In this work, we discuss the design and development of \emph{PromptAssist}, a prototype, accessible interface for creating T2I prompts. PromptAssist is a web-based application that allows users to create T2I prompts using a combination of keyboard and mouse input. PromptAssist uses a large language model (LLM) to suggest prompts based on whatever the user has input. PromptAssist offers three primary features for creating prompts:

\begin{enumerate}
\item automatically-generated suggestions for creating prompts or adding details to prompts;
\item an accessible text entry interface that supports both text entry and pointer-based interaction;
\item a wizard-based workflow that guides the user in creating prompts.
\end{enumerate}

PronptAssist was created by a diverse team that primarily includes individuals with motor disabilities (4 of 5 authors identify as having a motor disability); its design is motivated by our personal experiences interacting with inaccessible user interfaces and solving our own accessibility problems.

\section{Related Work}
\subsection{Text-to-Image Generation Models}
Recent years have seen a great expansion in the capabilities of T2I models, resulting in more detailed and realistic images. Key models include Reed et al.'s Generative Adversarial T2I Synthesis model~\cite{reed2016generative} which employs conditional generative adversarial networks (cGANs) to create images from text, the Imagen~\cite{saharia2022photorealistic}, a T2I diffusion model and Parti~\cite{yu2022scaling}, which uses sequence-to-sequence techniques and ViT-VQGAN tokenization to generate photorealistic images. 

Recently, much attention has been drawn to models that are available to users, often through public betas, include OpenAI's DALL-E 2~\cite{ramesh2022hierarchical}, which creates high-resolution images from text using 650 million image-text pairs, Midjourney, an open beta tool popular among artists available as a Discord chatbot, and Stable Diffusion~\cite{rombach2022high}, an open-source model operable on users' own devices. Research, like that by Chang et al.~\cite{chang2023prompt}, explored how users create and share prompts within online communities. Our work provides some early insights into the potential use of T2I models by individuals with disabilities.

\subsection{Improving Accessibility with AI}
Significant strides in AI have enhanced assistive technologies for individuals with motor disabilities, streamlining healthcare~\cite{sunarti2021artificial}, rehabilitation~\cite{guo2021artificial, aggogeri2019robotics, 10.1145/3373625.3416990}, education~\cite{zdravkova2022cutting, ranchal2013using}, and employment~\cite{fruchterman2018expanding, 10.1145/3531146.3533169, whittaker2019disability}, novel engagement modes in digital entertainment and video gaming~\cite{foley2012technology, orero2007accessible, taheri2021exploratory} through machine learning, computer vision, and natural language processing~\cite{10.1145/3373625.3416990, novak2015survey, lobo2014non, subasi2021ensemble, chavarriaga2017heading, yi2011design, taheri2021design}.
Notably, language model advances have revolutionized augmentative and alternative communication (AAC) options for non-verbal individuals~\cite{10.1145/3544548.3581560}, and systems like Voiceitt have amplified speech capabilities for those with severe speech impairments~\cite{murero2020artificial}. 

The art world has also seen a growing integration of AI to facilitate creation for artists with disabilities, such as AI working in tandem with prosthetics and bionics to improve accessibility~\cite{ikeda2022online, Zbiciak2023-ux, Zbiciak2023-mb, aldridge2021systematic, dazeddigitalCouldIncrease}. In this work, we focus on T2I tools—a groundbreaking AI-driven medium for artistic visual generation. By leveraging LLMs, we aim to boost interface accessibility, hence extending their utility to a wider audience. To our knowledge, the use of LLMs in visual creation via T2I models has not yet been explored.

\subsection{Accessibility and Creative Tools}
Computers, even general-purpose software can enable art creation for people with disabilities~\cite{creed2018assistive}. Researchers have explored innovative interfaces to facilitate creative tasks~\cite{ginsburg2020disability, rapoport2008engaging, flewitt2014touching}. Accessible creative tools emphasize simplified interfaces and designs catering to users with diverse abilities~\cite{burger1994improved, williams2012disability, mulfari2015computer}. 

Notably, work has been done to make music composition accessible with tools like AUMI~\cite{aumi}, Cyclops~\cite{payne2020cyclops}, EyeHarp~\cite{vamvakousis2016eyeharp}, and Hands-Free Music~\cite{microsoftMicrosoftHandsFree} using diverse inputs.
Several projects have facilitated visual art creation for those with motor disabilities. EyeDraw~\cite{hornof2003eyedraw}, EyeWriter~\cite{eyewriterEyeWriter} uses eye movements, and VoiceDraw~\cite{harada2007voicedraw}, uses non-speech vocalizations for drawing. Our work introduces an accessible visual art creation approach utilizing generative AI models. Unlike previous methods requiring learning, our system uses LLMs to generate creative text with reduced click or keystroke efforts, relying on existing interaction methods.

\section{PromptAssist: An Accessible Prompting Interface for Text-to-Image Models}

We developed PromptAssist as a way to explore accessibility challenges inherent in T2I tools, and potential ways to mitigate those challenges. The initial project idea was developed by three of the researchers (Taheri, Rostamzadeh, Kane) after several weeks of testing existing T2I models and user interfaces. After brainstorming various ideas for more accessible T2I tools, the team settled on the goal of enabling \emph{easier composition of text-to-image prompts} for users who experience difficulty in typing long texts (which includes several of the authors).

\begin{figure}
    \centering
    \begin{tabular}{c}
       \includegraphics[width=.9\textwidth]{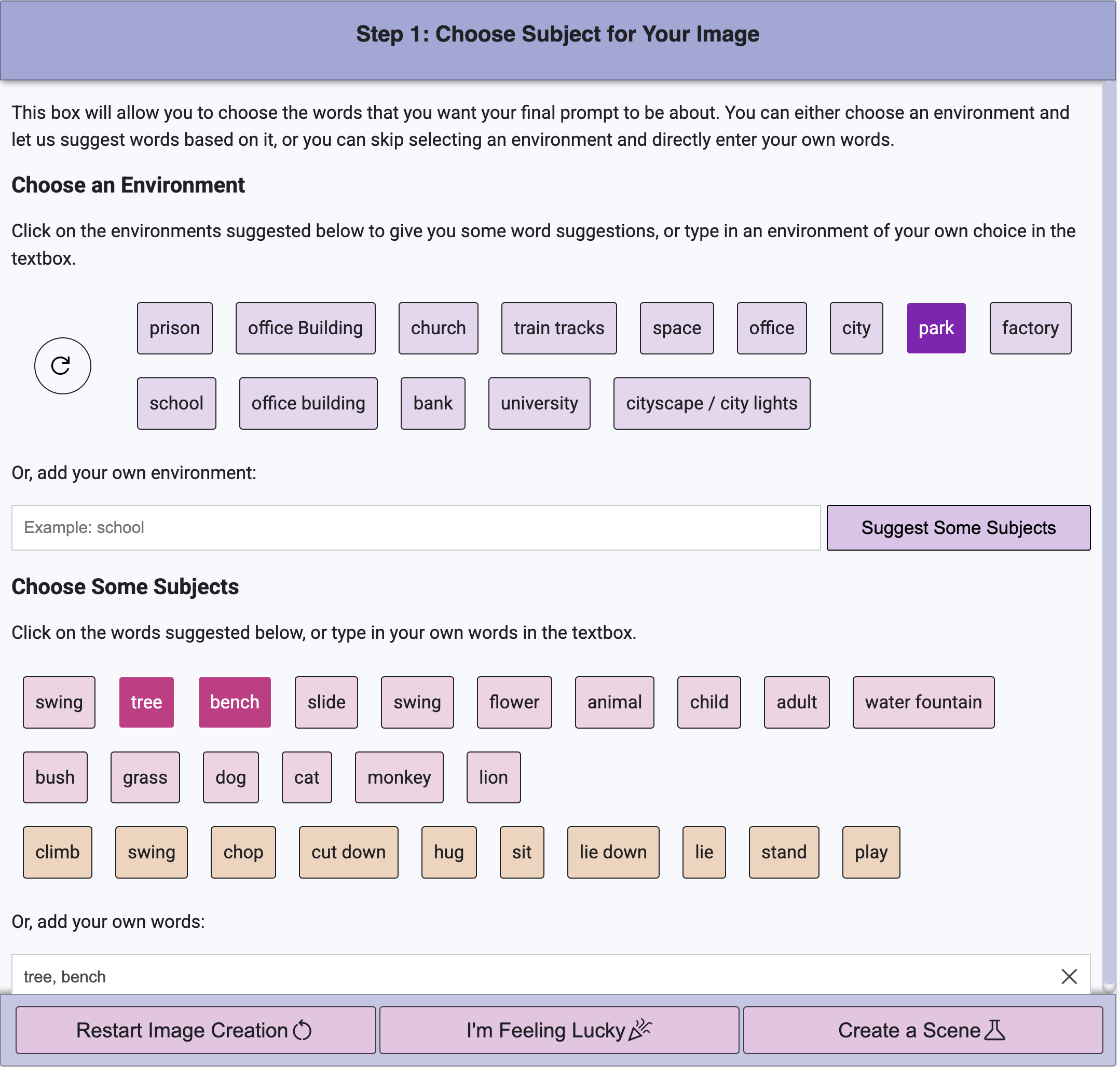}
         \\
         (A)
         \\\\
        \includegraphics[width=.9\textwidth]{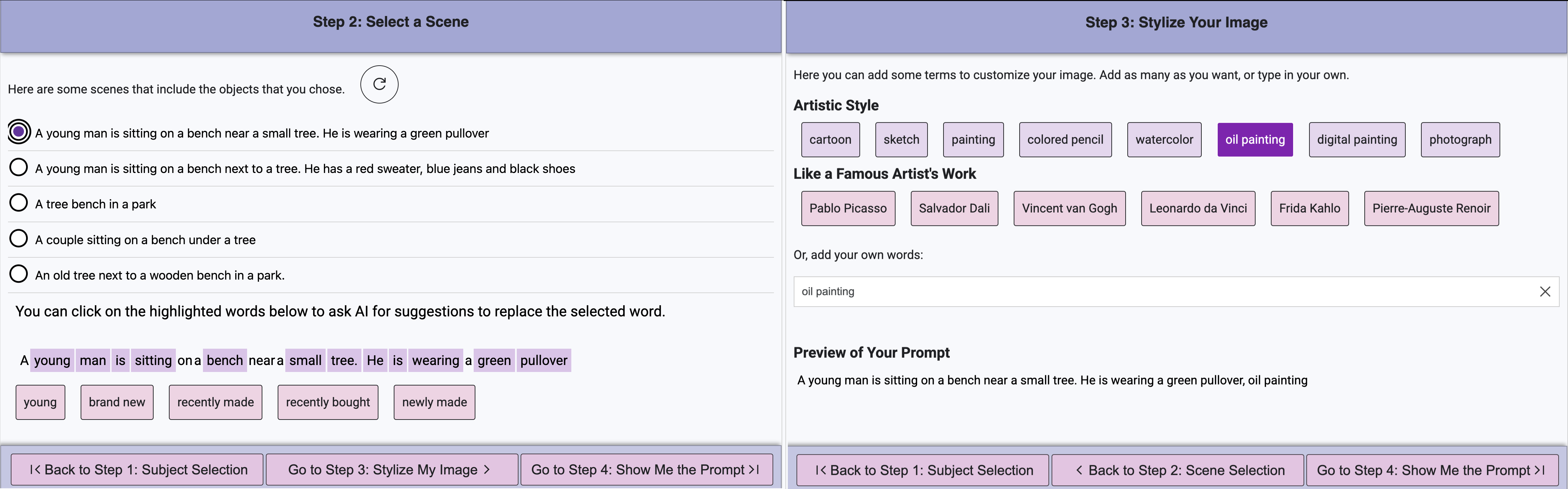}
        \\
        (B)
    \end{tabular}
    \caption{(A) Step 1 of PromptAssist allows users to select an environment and choose subjects and optionally actions within that environment. (B) In Step 2, PromptAssist enables users to choose from a variety of prompts (left image), and in Step 3, it provides the facility to incorporate artistic style data (right image).}
    \label{fig:demo}
\end{figure}

The PromptAssist prototype focuses on enabling users to write usable T2I prompts with minimal input. PromptAssist was partially inspired by existing tools like Promptomania's Prompt Builder~\cite{promptomania}, which helps users come up with creative ideas for prompts by suggesting image attributes that can be included in a prompt, such as style (\emph{e.g.,} pencil drawing or photograph), color palette (\emph{e.g.,} monochrome), or artist style (\emph{e.g.,} Frida Kahlo). Promptomania provides lists of these attributes, usually with thumbnails that show example output; users can include specific attributes by selecting them from menus. While Promptomania was designed to address creative challenges, its design may solve some accessibility challenges as well. With PromptAssist, our team attempted to build on this menu-based prompt creation workflow, with a specific focus on 1) enabling users to create prompts using a variety of input methods, and 2) enabling users to create prompts with minimal input.

\subsection{PromptAssist Workflow}
To support the design goals we identified, PromptAssist uses generative \emph{text} models (\emph{e.g.}, large language models)  to aid users in creating prompts for a text-to-image model. Recent accessibility research has explored how LLMs can help users compose text~\cite{51361, 10.1145/3544548.3581560, 10.1145/3544549.3573870} and create interactive prototypes~\cite{jiang2022promptmaker}; with PromptMaker, we extend that approach to creating text-to-image prompts. Instead of offering a preset list of prompt ideas, as Promptomania does, PromptAssist can extend whatever input the user provides. For example, a user might begin a prompt by typing the word ``beach'', and the system could add stylistic and other details (\emph{e.g.}, ''watercolor painting of a beach at sunset''). The PromptAssist prototype was built using an internal transformer-based large language model~\cite{thoppilan2022lamda} comparable to models such as GPT-3~\cite{brown2020language}\footnote{See Appendix~\ref{appendix:prompts} for the language model prompts used in our prototype system.}.

PromptAssist uses a wizard-based workflow to walk users through creating a prompt. At each step, the user can input their own text, request assistance from the model, or skip that step. PromptAssist breaks down prompt creation into four components: 

\begin{enumerate}
    \item \emph{Environment}. Choose an environment in which the image takes place;
    \item \emph{Subjects}. Identify particular objects to be included in the prompt;
    \item \emph{Actions}. Describe actions taking place within the scene;
    \item \emph{Scene}. Select a scene relevant to the chosen subjects and actions. Editing the scene may involve typing modifications directly or replacing selected words with suggested alternatives;
    \item \emph{Style}. Describe a medium (\emph{e.g.}, photograph) or other style characteristics.
\end{enumerate}

\begin{figure}[b!]
    \centering
    \begin{tabular}{c}
        \begin{minipage}{0.96\textwidth}
            \includegraphics[width=\textwidth]{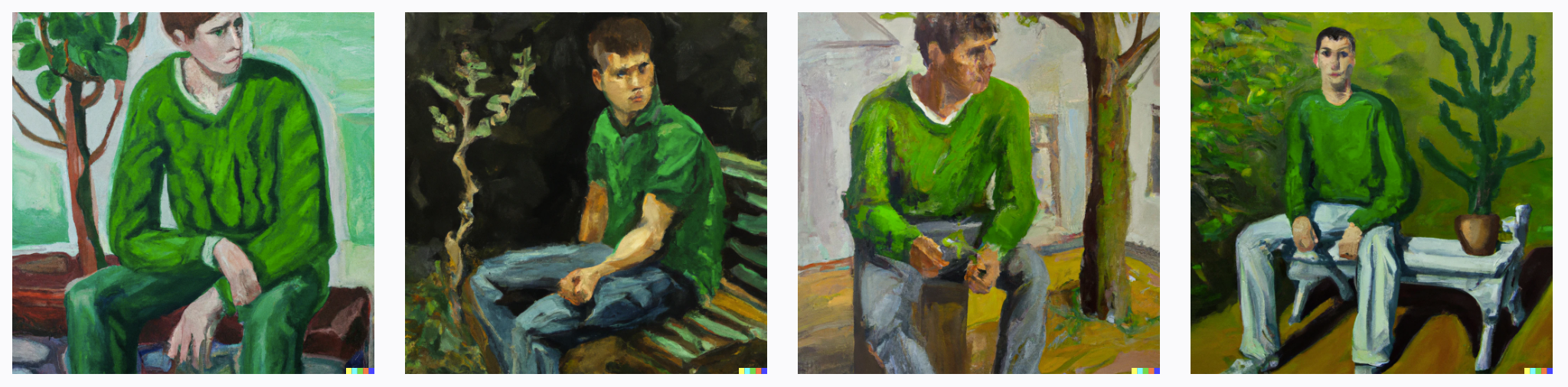}
            \caption{DALL-E 2 output for prompt ``A young man is sitting on a bench near a small tree. He is wearing a green pullover, oil painting.'' This prompt was created by the authors using suggestions from PromptAssist.}
            \label{fig:dalle}
        \end{minipage}
    \end{tabular}
\end{figure}

To support accessibility, information can be typed in at each step, or users can choose from suggestions using the mouse. At each step, the user can edit the suggested text or generate new suggestions. Figure~\ref{fig:demo} demonstrates an example interaction with PromptAssist. In the first step as depicted in Figure~\ref{fig:demo}:A, the user chooses an environment from a list of environments - in this case a \emph{park}. They then add some objects to the scene - a \emph{tree} and a \emph{bench}. Moving on to step 2, as illustrated in Figure~\ref{fig:demo}:B-left, the user selects a contextual scene from suggested ones - in this example \emph{``A young man is sitting on a bench near a small tree. He is wearing a green pullover''}. Lastly, as shown in Figure~\ref{fig:demo}:B-right, the user adds an artistic style - here \emph{oil painting} to the scene. The resulting prompt was copied from PromptAssist into DALL-E and the resulting images are shown in Figure~\ref{fig:dalle}.

\section{Research and Co-Design of PromptAssist}
PromptAssist has been developed through iterative development over the course of several months. The PromptAssist team contains multiple researchers with motor disabilities who themselves are users of the system.

\subsection{Research Team and Process}
Our multidisciplinary research team led by primary investigators, Taheri, Rostamzadeh, and Kane, was composed of experts from computer science, human-computer interaction, accessibility, and computer vision fields. Taheri, who superheaded the development of PromptAssist and led the paper writing process, has personal experience living with Spinal Muscular Atrophy (SMA). Being restricted to using her thumb on a touchpad with a typing speed to an average of 13 words per minute, served as a significant motivator for this project. Her experiences, coupled with the hurdles faced by similarly situated individuals using creative tools and her computer engineering background, inspired this endeavor towards more accessible technology. This personal connection further invigorated our team's enthusiasm for crafting an accessible T2I interface.

Team members contributed to research sessions, feedback, and project outputs. The team utilized Google Chat and Google Meet for real-time discussion and Google Docs for collective note-taking and resource exchange to maintain efficient communication.

\subsection{Iterative Development of PromptAssist}
Taheri, Rostamzadeh, and Kane conceived the initial concept for PromptAssist, and developed its early iterations. They refined this prototype through testing and recurring group discussion. Both Taheri and Kane used their own personal accessibility tools during testing. Once the PromptAssist prototype matured that could be more easily tested, they recruited Izadi and Shriram via an internal disability interest group within their organization. While Izadi and Shriram did not have a background in research, they joined the project because of their interest in shaping development of more accessible T2I technologies and to gain experience in accessibility research. 

Taheri, Izadi, Shriram, and Kane were present at every test session, while Taheri generally served as the session facilitator. The research team tested prototype of PromptAssist on their own devices, but generally tested their prompts using an internal version of Parti~\cite{yu2022scaling}, an autoregressive text-to-image model that produces images comparable to models such as DALL-E (although it is worth noting that prompts generated by PromptAssist can be copied and pasted into any T2I model).

\subsection{Test Session 1: Collaborative Planning and Discussion}
This session functioned as an informative gathering. Taheri detailed the research goals and future plans, particularly, guiding less experienced members, Izadi and Shriram, about their roles in iterative testing and feedback. A group chat room was created for the research duration to facilitate questions and image sharing among the team.

\subsection{Test Session 2: Experimenting with Existing Text-to-Image Tools}
In this session, the team lead familiarized the team with existing T2I tools, encouraging them to experiment with the current Parti user interface, which consisted of only a text box and submit button. The team explored the tools' potential and boundaries by generating images based on varying concepts, styles, and detail levels. Izadi, for example, focused on an idea of a bird with a corgi's face flying over San Francisco. Shriram experimented with multiple prompts to understand the safety filter's operation. Kane used different color terms to observe their impact on cat portrait images. The team members communicated their experiences throughout the session in a ``think-aloud'' format, and Taheri and Kane documented the proceedings.

\subsubsection{Feedback on Current T2I Systems}
The team members, particularly those new to T2I systems, were impressed by the output quality but agreed on the need for usability and accessibility improvements. Issues identified included the absence of access keys or keyboard shortcuts, the inability to cancel the slow image generation process, and the lack of autocomplete or grammar correction. The possibility of benefiting from pre-made prompt categories or ideas was also discussed. One theme that arose during testing was difficulty in generating longer prompts or iterating on prompts. Shriram, in particular, found it difficult to type longer or more prompts. The team discussed how to refine prompts using natural language instructions like ``make this more descriptive.'' Izadi expressed interest in using natural language to edit or merge images generated by the model.

\subsection{Test Session 3: Experimenting with PromptAssist}
In this collaborative session led by Taheri, the team tested the PromptAssist prototype. Taheri explained how PromptAssist addressed the usability and accessibility issues identified in the previous T2I systems session. Team members worked together and shared feedback via a ``think-aloud''.

This version of PromptAssist, while similar to the one described above, required users to compose their prompt in a specific order, without skipping steps.

\subsubsection{Feedback from the session}

Shriram spent much of his time generating images related to the ocean. He was particularly interested in what subjects were added to the image, noting that requesting ocean images tended to include plants but not animals. Izadi noted that the user interface did not quite match their goals; while they appreciated that PromptAssist offered suggestions, they wanted to be able to start with their own specific prompt and build upon it. While Izadi noted that selecting parts of prompts using the mouse reduces the need for typing, they desired the ability to do both. 

The team suggested a variety of user interface changes, including: reducing white space in the interface; increasing color contrast; and making it easier to go back or restart a prompt. This early version usually provided 3-5 suggestions in each category; participants requested that ten or more suggestions would be useful, especially when coming up with their initial idea for the prompt. As discussed in the first session, adding autocomplete would be helpful so that users could enter a partial prompt and let the model finish it.

\subsection{Test Session 4: Revised PromptAssist}
Following the previous session, Taheri updated PromptAssist per user feedback. Enhancements included: adding a multi-page layout, improving color contrast, writing human-readable error messages, increasing the number of suggestions, allowing users to generate additional suggestions, adding the option to skip steps like artistic style, and enabling keyboard-only navigation of all buttons and menus.

As in the previous session, Taheri walked the team through the updates made in response to their feedback. The team members tested the enhanced PromptAssist, sharing their thoughts. They all found that the changes made it easier to navigate through the process and to follow their own creative ideas. The revised prototype's notable advancement was its flexibility in allowing user-defined creative processes over prescribed sequences. In contrast to the prior version, which enforced a specific prompt format, the allowed users to enter their own prompts, accessing assistance only when needed.

\section{Discussion}

\subsection{Improving Usability and Accessibility of T2I Tools}
This paper documents the co-design and evolution of an accessible prompt creation tool for T2I systems, which was designed in response to feedback from disabled team members who found typing prompts, particularly longer ones, challenging in the existing text-based interface, which was essentially just an HTML text input control.

The creation of PromptAssist provides ways to enhance T2I interface accessibility. It facilitates prompt creation through typing, pointing, clicking, or a combination, adhering to the WCAG's concept of \emph{operability}—allowing any input device to perform all actions. Unlike existing tools that limit input to predefined categories, PromptAssist uses an LLM to offer contextual suggestions, allowing users to expand their own ideas, rather than choosing from a preset list of image ideas. In addition to enabling more accessible forms of input, these contextual suggestions could be beneficial for all users regardless of their abilities.

\subsection{Unleashing Creativity with T2I Tools}
It is clear that T2I models can enhance the creative abilities of people, enabling them to create images that they would be unable to create by other means. T2I systems may be especially empowering to individuals who are unable to effectively use other image creation tools because of accessibility barriers. 
The focus on text input in current T2I models is beneficial in some ways, as many people with disabilities have already found accessible ways to input text. However, as shown in this work, there remains the opportunity to increase accessibility and ease of use of these systems. 

\subsection{Creativity \emph{vs.} Ease of Use}
One tension that arose in designing PromptAssist's user interface was  balancing creative flexibility and ease of use. 
While pre-generated contextual prompt ideas can help users who find typing challenging, it risks limiting creativity and user autonomy. Interviews with experienced T2I users revealed that prompt crafting is viewed as part of their creative work~\cite{chang2023prompt}. Over-reliance on language generation could also diminish the perceived independence of a user with disability, giving the impression that the system, rather than the user, is doing the work~\cite{10.1145/3544548.3581560}.

To mitigate concerns about creativity and autonomy, PromptAssist enables users to view system suggestions, which they can then accept, modify, or reject. This approach amplifies their original ideas without making creative choices for them, an aspect that remains crucial in the development of accessible creative tools.

\section{Concluding Remarks and Future Work}
Our iterative development of PromptAssist resulted in a prototype that supported accessible input while maintaining creative flexibility. The system, powered by an LLM, introduces contextual suggestions and a robust interface to ease prompt creation and improve user experience. However, for the future work, it could further benefit from multimodal input capabilities such as typing, pointing together with speech, and body movements for prompt creation, along with the ability to adjust the prompt based on previously generated images.

Current key features of PromptAssist (\emph{e.g.,} contextual suggestions,) can notably enhance mainstream T2I systems, and other tools for creative writing or video creation. Its potential extends to aiding users with sensory or cognitive disabilities, and offering expert users more customization such as complex prompt creation by expert users.  

Consequently, Generative AI models, including T2I models, with the aid of user-friendly interfaces, can revolutionize media creation by overcoming accessibility challenges and supporting individuals of all abilities in their creative pursuits. Future work should focus on these enhancements to further the reach and utility of such tools. Additionally, fostering a platform for users to share their creations, collaborate on art projects, and exchange ideas, may not only improve user experience but also promote artistic growth and unique styles stemming from user and model collaborations.

\bibliographystyle{ACM-Reference-Format}
\bibliography{sample-authordraft}


\pagebreak
\appendix

\section{LLM Prompts}\label{appendix:prompts}
This section includes the prompts that were used to generate suggestions in the PromptAssist prototype. These prompts evolved as the prototype developed; this appendix includes the prompts used in the final round of testing. These prompts use a system inspired by PromptMaker~\cite{jiang2022promptmaker}; the prompt is passed to a large language model, and the output is returned to PromptAssist. In these examples, variable input to the prompt is represented by the symbol \textbf{<input>} and output from the language model is represented by the symbol \textbf{<output>}.

\subsection{Suggest Environments}

\begin{verbatim}
Name: environment
Suggestion: university
Name: environment
Suggestion: ocean
Name: environment
Suggestion: hospital
Name: environment
Suggestion: <output>
\end{verbatim}

\subsection{Suggest Subjects for an Environment}

\begin{verbatim}
Environment: school
Suggestions: blackboard, teacher, chair, book, student, class, eraser, whiteboard, notebook, pen, pencil, eraser, 
paper
Environment: work office
Suggestions: desk, computer, pen, paper sheet, folder, fax, phone, pencil, paper shredder, light
Environment: forest
Suggestions: tree, animal, bird, monkey, lion, fox, eagle, plant, flower, insect, tiger, horse, wolf
Environment: home
Suggestions: TV, bed, table, sofa, couch, light, console table, remote control, carpet, rug, room, kitchen
Environment: sea
Suggestions: fish, jelly fish, star fish, shark, dolphin, whale, island, boat, ship, coral, crab
Environment: <input>
Suggestions: <output>    
\end{verbatim}

\subsection{Suggest Actions for Subjects}

\begin{verbatim}
word: tv
verbs: watch, work, fix, turn on, turn off, turn up, turn down, put, pick up
word: cat
verbs: play, eat, sit, run jump, scratch, pet, sleep, feed, jump, meow, brush, groom, bathe, cuddle, love
word: pool
verbs: fill, swim, drawn, go down, dive, jump, splash, play, go down, drink, eat, throw, throw up, spit, pee, pee in
word: paper
verbs: write, read, draw, cut, tear, color, crumple, make, throw, pick up, put down, fold, take, give, put away, color, spin
word: <input>
verbs: <output>    
\end{verbatim}

\subsection{Generate Prompts from Scenes and Subjects}

\begin{verbatim}
words: dog
scene: A small dachshund doing a kickflip on a skateboard
words: cat, tree
scene: A young DSH cat sitting on a small tree branch with leaves in a garden
words: space, rocket
scene: A black and white space ship with a rocket attached to the engine. There is a trail of smoke following the 
 rocket flying through the space full of planets and stars
words: chair, cup, parrot
scene: A parrot with a blue feather and a black/grey beak sitting on a high-backed chair. There is a small table 
 next to the chair and a red cup of tea on the table
words: fish, waves
scene: A red snapper fish with a white body and black eyes and mouth is swimming through the waves at the ocean
words: flying car, fly, bird
scene: a flying car is driving in the sky and a bird is flying next to the car
words: paper sheet, fold, folder
scene: a paper sheet is being folded on a blue folder with a white paperclip on it
words: sofa, sleep
scene: A man is sleeping on the sofa. He has a red t-shirt, blue shorts and brown hair,
words: monkey, swing, tree
scene: A small monkey is swinging on a tree branch. There is a red banana next to the branch
words: take blood from
scene: A doctor is taking blood from a syringe and putting it in a small tube,
words: eiffel tower, old man, old woman
scene: An old man and an old woman are drinking wine next to the eiffel tower.,
words: cat, mouse
scene: An outdoor cat patiently waiting by a mouse hole for its next meal.
words: <input>
scene: <output>   
\end{verbatim}

\subsection{Generate Synonyms for Words}
\begin{verbatim}
word: blue
replacements red, pink, orange, yellow, purple, green, brown
word: small
replacements big, tiny, giant, medium, large, huge, miniature
word: young
replacements old, adult, child, teenager, infant, baby, middle-aged
word: <input>
replacements <output>   
\end{verbatim}

\end{document}